\documentclass[10pt, conference, compsocconf]{IEEEtran}
\usepackage[utf8]{inputenc}
\usepackage[hidelinks]{hyperref}
\usepackage[T1]{fontenc}
\usepackage{graphicx}
\usepackage{amssymb}
\usepackage{tabularx}
\usepackage{url}
\usepackage{multirow}
\usepackage{sparklines}
\usepackage{float}
\usepackage{multibib}

\usepackage{authblk}

\usepackage{booktabs}
\usepackage{array}
\newcolumntype{v}[1]{>{\raggedright \hspace {0pt}}p{#1}}
\usepackage{colortbl}

\usepackage{booktabs}
\usepackage{array}
\usepackage{colortbl}
\usepackage{multirow}
\usepackage{longtable}

\newcolumntype{v}[1]{>{\raggedright \hspace {0pt}}p{#1}}
\newcolumntype{G}[1]{>{\columncolor{gray90}}#1}

\definecolor{Gray}{gray}{0.8}
\definecolor{gray25}{gray}{0.25}
\definecolor{gray50}{gray}{0.50}
\definecolor{gray75}{gray}{0.75}
\definecolor{gray90}{gray}{0.9}

\newcommand{\grayrow}{\rowcolor{gray90}}

\title{Safety-Critical Systems and Agile Development: \\ A Mapping Study}
\author[1,2]{Rashidah Kasauli}
\author[1]{Eric Knauss} 
\author[2]{Benjamin Kanagwa}
\author[1] {Agneta Nilsson}
\author[1] {Gul Calikli}
\affil[1]
{Chalmers $\mid$ University of Gothenburg, Sweden}
\affil[1]{\textit{{rashida@chalmers.se},\{eric.knauss, agneta.nilsson, gul.calikli\}@cse.gu.se}}
\affil[2]{Makerere University, Uganda}
\affil[2]{\textit{\{rnamisanvu, bkanagwa\}@cis.mak.ac.ug}
}

\newcites{pri,sec}{Primary Papers for the study, References}

\date{today}
\begin{document}
\maketitle

\begin{abstract}
In the last decades, agile methods had a huge impact on how software is developed.
In many cases, this has led to significant benefits, such as quality and speed of software deliveries to customers. 
However, safety-critical systems have widely been dismissed from benefiting from agile methods.
Products that include safety critical aspects are therefore faced with a situation in which the development of safety-critical parts can significantly limit the potential speed-up through agile methods, for the full product, but also in the non-safety critical parts.
For such products, the ability to develop safety-critical software in an agile way will generate a competitive advantage.  
In order to enable future research in this important area, we present in this paper a mapping of the current state of practice based on {a mixed method approach}.
Starting from a workshop with experts from six large Swedish product development companies we develop a lens for our analysis.
We then present a systematic mapping study on safety-critical systems and agile development through this lens in order to map potential benefits, challenges, and solution candidates for guiding future research.
%
\end{abstract}

\begin{IEEEkeywords}
Safety-critical systems, agile, continuous integration, continuous delivery, continuous deployment, systematic mapping study
\end{IEEEkeywords}

\section{Introduction}
A system is safety-critical if its failure can cause financial loss, damage to the environment, injury to people and in some cases, loss of lives \citesec{laplante2017software,hatcliff2014certifiably}.
The use of software in safety-critical systems has 
{continuously increased to an} extent 
{where} software failures can impair system safety. 
Examples in the automotive domain include software in a break system, which in case of failure could result in unacceptable hazards, but also active safety functions that override driver behavior in certain situations to avoid a crash. 
{Similar examples can be found in o}ther domains (e.g. railway \citesec{provenzano2017specifying}, avionics \citesec{grant2016requirements})

{Safety-critical systems (SCS) are heavily regulated and require certification against industry standards by the relevant governing body \citesec{vuori2011agile}. 
For that reason, they have to be designed with safety in mind. 
However, as opposed to the hardware where general safety design principles have been incorporated into standards, the standards for development of safe software {are still developing}. 
These standards, for instance ISO26262 for automotive, IEC61513 for nuclear, IEC62304 for medical and DO-178B for avionics %
{aim to ensure that best engineering practice is followed and are often based on traditional, plan-driven approaches \citepri{notander2013}\footnote{In this paper, we use numbered references for regular citations and abbreviated style for primary papers of the mapping study.}. This is related to the common criticism that agile methods neglect upfront analysis with a focus on scenarios for describing requirements \citesec{Meyer2014}. Because the standards are written like that and because people think that agile methods are not suitable, SCS development is most often characterized by highly-structured, plan-driven approaches such as V-model or waterfall \citesec{ge2010iterative}.

While agile methods have significantly changed the way most software is developed, safety-critical software was widely excluded from this trend \citesec{Meyer2014}.
Arguments for this exclusion relate to the need for documenting safety cases and the need to analyze and specify safety requirements upfront, both thought to be in conflict with agile values.
Thus, companies have different approaches to develop software, depending on whether it is  safety-critical or not (e.g. allowing faster time to market of non-critical software such as infotainment) \citesec{Kuhrmann2017}. 
Recently, however, the situation has changed through potentially disruptive trends that significantly increase the need for short development cycles and quick time to market. 
Complex dynamic systems-of-systems require components of different vendors to interact at runtime and the dynamic nature of such environments can make continuous deployment a necessity for product success and for maintaining functional safety, e.g. in the automotive domain \citesec{Pelliccione2017}. 
The huge attention given to autonomous driving and intelligent vehicles has lead to a jump in complexity of SCS.
Companies like Tesla have demonstrated how the ability to deploy new functions and explore their performance in the field can yield huge advantages.
Software companies such as Google, Apple, and Amazon have pushed in the automotive market, increasing the 
{need} 
to develop competencies in continuous software engineering.
Slowly, companies developing SCS realize the competitive advantages that agility can provide \citepri{cawley2010lean}.


Existing reviews on agile development and safety \citesec{kaisti2013agile,cawley2010lean} mostly focus on determining whether agile methods can improve embedded systems development \cite{kaisti2013agile} in general and on how these agile methods are being adopted in these environments. Kaisti et al.'s study \cite{kaisti2013agile} does not particularly focus on 
safety-critical systems whereas Cawley et al. \cite{cawley2010lean}, with a focus on regulated environments, found few works indicating a low but growing adoption of agile methods. 
Thus, it is about time to investigate  the state of supporting agile and continuous development of safety-critical software.

In order to enable future research in this important area, we present in this paper a mapping of the current state of practice based on {a mixed method approach. }
Starting from a workshop with experts from six large Swedish product development companies, we develop a lens for our analysis.
The aim of this mapping study is to evaluate and present the combination of agile development methods and SCS.
Our goal is to provide a map that can enable  future research and method development for researchers and practitioners. Specifically, we map suggested benefits, challenges, and solution candidates as well as their relation to up-front analysis, just-in-time activities of agile teams, and infrastructure to support these. 
We aim to answer the following research questions. 
\newcommand{\researchquestionone}{
{What research exists about agile development {of} SCS?}}
\newcommand{\researchquestiontwo}{What are the key benefits of {applying} agile {methods and practices} in SCS development?}
\newcommand{\researchquestionthree}{What challenges exist with agile development of SCS? }
\newcommand{\researchquestionfour}{What solution candidates (e.g. principles and practices) promise to address  challenges with respect to agile development of SCS?}

\noindent \emph{\textbf{RQ1:} \researchquestionone}

\noindent \emph{\textbf{RQ2:} \researchquestiontwo}

\noindent \emph{\textbf{RQ3:} \researchquestionthree}

\noindent \emph{\textbf{RQ4:} \researchquestionfour}

Thus, this paper contributes an overview of existing research, potential benefits, challenges, and solution candidates of agile development of SCS.  
We discuss our findings in three dimensions: The necessity of up-front analysis, the ability to shift effort into just-in-time activities, and the potential of better infrastructure to support these. 
We expect this contribution to be of value for practitioners aiming to optimize their processes as well as for researchers who may base future research on our synthesis.

\section{Methodology}
In order to provide the required context while answering our research questions, two complimentary activities have been used; a cross-company workshop and a mapping study. This section provides an elaboration on these two processes.

\textbf{The Cross-Company Workshop:}
We started to get interested in the interplay of agile methods and SCS based on ongoing collaborations and research projects on continuous integration and deployment. 
Through this existing network, we invited experts responsible for adoption of continuous integration, continuous deployment, and agile methods as well as safety experts from large Swedish companies to a workshop on SCS in continuous software engineering.
During this workshop, representatives from six companies attended. 
We presented a very preliminary screening of literature, followed by presentations from industry participants.
We then had an open discussion through which we arranged the information presented and agreed on critical aspects to be further investigated. 
From this activity, we derived a lens for conducting our mapping study {to guide our research}. 
We use this lens to provide an overview of research in the field, specifically focusing on identifying \emph{proposed benefits}, \emph{challenges}, and \emph{potential solutions} and to discuss these findings based on their relevance to \emph{upfront analysis} (i.e. activities to be done before agile teams can continuously add value through agile development methods), \emph{just-in-time activities} (i.e. activities done during agile development sprints), and \emph{long lasting infrastructure beyond a single project}. 



\textbf{The mapping study:}
The key steps in the mapping study followed guidelines by Petersen et al. \citesec{petersen2008systematic} while also borrowing some concepts from the established method of systematic literature reviews \citesec{kitchenham2004}.  
Kitchenham and Charters \citesec{kitchenham} outlined three phases of performing systematic reviews, i.e., planning, conducting and reporting the review. 
These phases were later simplified by Pertersen et al. \citesec{petersen2008systematic} into five stages that include: defining research questions (see Section 1), conducting a search for primary studies, screening primary studies based on inclusion/exclusion criteria, classifying the primary studies, and data extraction and aggregation. 
These are explained in the subsections that follow.
\begin{table}[t]
\caption{Number of papers found, excluded and included 
}
\label{tab:search-results}
\centering
\resizebox{\columnwidth}{!}{%
\begin{tabular}{v{0.5\columnwidth}rr}
\toprule

\multirow{2}{*} {{\textbf{Inclusion/exclusion criterion}}} & \multicolumn{2}{l}{\textbf{Number of documents}} \tabularnewline \cmidrule{2-3}
&\bf Excluded& \bf Included\tabularnewline\midrule

Search from 2001 to 2017 & & 1986 \tabularnewline
\grayrow Limit results to subject \emph{Computer Science} in Scopus &$-1420$ &  546\tabularnewline
Exclude documents without authors & $-81$ & 465\tabularnewline
\grayrow Exclude non-english documents & $-6$ & {458}\tabularnewline \midrule

{\bf{Exclusion based on title and abstract}} & $-391$& 69 \tabularnewline
\multicolumn{3}{l}
{\em - Document mentions safety but not in the agile development context.}\tabularnewline
\multicolumn{3}{l}
{\em - Document discusses agile without a particular focus on SCS.}\tabularnewline
\multicolumn{3}{l}
{\em -  The subject area was not software engineering or development.}\tabularnewline
\multicolumn{3}{l}
{\em - The document is not peer reviewed.}\tabularnewline \midrule
\grayrow {\bf{Criteria based on full text}} & $-35$ & \textbf{\underline{$\mathbf{33+1}$}}\tabularnewline
\grayrow \multicolumn{3}{l}
{\em - Exclude documents where full text could not be obtained.}\tabularnewline
\grayrow \multicolumn{3}{l}
{\em - Exclude documents to which an extended version was found and could be included.}\tabularnewline
\grayrow \multicolumn{3}{l}
{\em - Exclude documents that do not allow extraction of context information.}\tabularnewline
\grayrow \multicolumn{3}{l}
{\em - Include documents presenting an experience report.}\tabularnewline
\grayrow \multicolumn{3}{l}
{\em - Include documents presenting a technical solution paper or an experiment.}\tabularnewline
\grayrow \multicolumn{3}{l}
{\em - Include documents presenting a case study with students or practitioners.}\tabularnewline
\grayrow \multicolumn{3}{l}
{\em - Include other documents (e.g. short papers) that include industry experience.}\tabularnewline
\grayrow \multicolumn{3}{l}{\em - Include papers obtained by snowballing given they meet the above criteria.}\tabularnewline
\bottomrule
\end{tabular}}
\end{table}
%

\subsection{Search strategy}
Basing on our research questions, we selected the major search terms; "safety-critical systems" and "agile development". We then {adapted the} PICO (Population, Intervention, Comparison and Outcome) criteria proposed by Kitchenham and Charters \citesec{kitchenham} to construct the search terms used in {our} final search. 
{With respect to} \emph{population}{, we } include terms {related to} safety-critical systems. These are synonyms or terms related to the major term and include safety-critical software, safety, safety critical, regulated, regulation or software intensive.
{With respect to }intervention, we included terms relating to agile development and agile methods. These include agile, agility, continuous delivery, continuous deployment, continuous integration, scrum, xp, agile method, agile process and agile practices.
{Specifically}, we {relied on a literature review on agile methods \citesec{diebold2014agile} to select most commonly used search terms.} 
We did not use the term `software development' {in our search} since it was too specific and limiting, even though it is the context in which we are working. 
We do not aim to compare methods and therefore did not use the comparison criteria nor did we use the outcomes criteria since the scope of the outcomes of this study is hard to limit. 
A pilot literature search was then performed to verify whether omission or addition of one or more search terms could lead to a decrease or increase in the documents returned. 
Search terms that did not add any new documents were dropped and others kept. 
After several iterations, the following search string was defined: 

\begin{quote}
\scriptsize{\textit{TITLE-ABS-KEY(agile OR agility OR "continuous integration" OR "continuous deployment" OR "continuous delivery" OR scrum OR "agile practices" OR xp) AND TITLE-ABS-KEY("safety-critical systems"  OR  "safety-critical software" OR  safety OR "safety critical" OR regulated OR regulation OR "software intensive")}}
\end{quote}


\subsection{Inclusion and Exclusion criteria}
Using the search string obtained, we performed an automated search on Elsevier Scopus (www.scopus.com), a database indexing many relevant venues and journals. The search returned 1986 papers which were then limited to the 'Computer Science' subject in Scopus leaving 546 papers. From these, all duplicates, editorials and panel paper were removed leaving 465 papers.
The papers that were included were those that presented any kind of empirical study dealing with application of agile methods in a safety-critical environment, which were written in English and which were published from 2001, when the agile manifesto was launched, to 2017.

On the other hand, studies were excluded if they were pure discussion and opinion papers, duplicates (e.g. conference paper extended into journal), {or} if their main focus was not agile development in SCS. 
A summary of the selection criteria and as well as the percentage of the paper excluded and included based on our criteria are shown in Table \ref{tab:search-results}. 

After removing the 6 non-english papers, the first author went through the titles and abstracts of all  remaining 458 papers to determine their relevance for the study. 
The titles and abstracts were taken together since from the preliminary search, we realised that some studies' titles are not that explicit and could lead to elimination of relevant studies. 
In this step 391 papers were excluded, since their title and abstract were clearly not about agile software development and SCS.
In order to increase reliability of our study, we had a random sample of 20 papers rated by three of the other authors and computed an inter-rater agreement ({Fleiss Kappa $\kappa = 0.595$}). 
All disagreements were discussed and resolved among the researchers, leading to an agreed result of 69 papers.
Then, the first two authors went through the full texts of these remaining 69 papers collaboratively and selected 33 papers based on the inclusion and exclusion criteria in Table \ref{tab:search-results}, to which we added one more paper identified through snowballing \citesec{wohlin2014guidelines} for consideration (indicated with $+1$ in Table \ref{tab:search-results}),  resulting in a total of 34 papers.
\subsection{Data extraction and Synthesis}
With respect to our research questions, we then extracted metadata, practices used, principles followed, challenges faced, {and} {(potential)} benefits from the 34 selected papers based on a predefined template for data extraction presented in Table \ref{tab:extraction-template}.
\begin{table}
\centering
\caption{Table: Data extraction template}
\label{tab:extraction-template}
\begin{tabular}{lv{0.7\columnwidth}}
\toprule
\grayrow Data item& Description\tabularnewline\midrule
Title & The title of the paper. \tabularnewline
Authors & The full names of the authors.\tabularnewline
Year & Year of publication.\tabularnewline
Venue & Name of publication venue.\tabularnewline
Publ. type &Journal, conference or workshop.\tabularnewline
Summary & The main content of the paper.\tabularnewline
Benefits & The benefits of agile development in SCS.\tabularnewline
Challenges & Problems faced with agile development in SCS. \tabularnewline
Practices  & The practices used in that safety + agile environment.\tabularnewline
Principles  & The principles they follow if any.\tabularnewline
\bottomrule
\end{tabular}
\end{table}
This template allowed us to extract all the information we need for analysis into a MS Excel spreadsheet. The data obtained was analyzed both quantitatively and qualitatively. We used the themes agreed during the workshop with company experts, while basing a thematic approach, to group the results obtained into the upfront, just-in-time and infrastructure.
%
For synthesis, the metadata was summarised and proposed benefits, challenges and potential solutions were collated, analysed{, and categorized among the researchers.}



\subsection{Limitations and Threats to Validity}
Our literature search was limited to the Elsevier Scopus database. Even though Elsevier claims that Scopus is the most comprehensive database of peer-reviewed research abstracts\footnote{http://www.elsevier.com/elsevier-products/scopus}, additional databases could have provided more relevant studies. However, this was mitigated by snowballing, where the citations of the included papers 
were reviewed for consideration, leading to one additional paper.  
In order to mitigate potential misinterpretations of papers, we relied on more than one researcher in each key step of our research. 
Where appropriate, we calculate inter-rater agreement. 
Despite these efforts, we may have overlooked relevant papers or misinterpreted them in our synthesis. 
We aimed for a transparent description of our method to allow for recovery if this should be the case.

\section{Findings}
\newcommand{\slallspike}[0]{
\definecolor{sparkrectanglecolor}{named}{cyan}
\begin{sparkline}{12}
\sparkspike 0 0.6
\sparkspike .08 0.9
\sparkspike .16 0.1
\sparkspike .24 0.4
\sparkspike .32 0.4
\sparkspike .4 0.3
\sparkspike .48 0.2
\sparkspike .56 0.2
\sparkspike .64 0.1
\sparkspike .72 0.1
\sparkspike .8 0
\sparkspike .88 0.1
\definecolor{sparkspikecolor}{named}{red}
\end{sparkline}}

\newcommand{\slall}[0]{%
\definecolor{sparkrectanglecolor}{gray}{0.9}%
\begin{sparkline}{6}%
\sparkrectangle 0 1
\spark 0 0.1  .083 0   .167 0.1 .25 0.1 .33 0.2  .416 0.2  .499 0.3  .588 0.4 .671 0.4  .754 0.1  .837 0.9  1 0.5 /
\end{sparkline}}

\newcommand{\sljrnl}[0]{%
\begin{sparkline}{2}
\sparkrectangle 0 1
\definecolor{sparkspikecolor}{named}{red}
\sparkspike 0.5 0.5
\end{sparkline}}

\newcommand{\slmag}[0]{%
\begin{sparkline}{3}
\sparkrectangle 0 1
\definecolor{sparkspikecolor}{named}{red}
\sparkspike 0.25 0.5
\definecolor{sparkspikecolor}{named}{blue}
\sparkspike 0.75 0.1
\end{sparkline}}

\newcommand{\slconf}[0]{%
\begin{sparkline}{4}
\sparkrectangle 0 1
\definecolor{sparkspikecolor}{named}{red}
\sparkspike 0.2 0.5
\definecolor{sparkspikecolor}{named}{blue}
\sparkspike 0.5 0.1
\definecolor{sparkspikecolor}{named}{purple}
\sparkspike 0.8 2.2
\end{sparkline}}

\newcommand{\slws}[0]{%
\begin{sparkline}{4}
\sparkrectangle 0 1
\definecolor{sparkspikecolor}{named}{red}
\sparkspike 0.2 0.25 
\definecolor{sparkspikecolor}{named}{blue}
\sparkspike 0.4 0.01
\definecolor{sparkspikecolor}{named}{purple}
\sparkspike 0.6 1.1
\definecolor{sparkspikecolor}{named}{magenta}
\sparkspike 0.8 0.35
\end{sparkline}}

\subsection{RQ1: Existing research about agile development of SCS}

{Based on our search terms in combination with the defined inclusion and exclusion criteria, we selected {34 }papers}\footnote{{All papers:} {\tiny \url{https://docs.google.com/spreadsheets/d/19s4aTpB0Yy38mnYUh1F1TiQqPZEdHD1ZYP8VdC2QQdE/edit?usp=sharing}}}. 
{As shown in Table \ref{tab:papers}, we did not select any paper before 2006 and there has been a \emph{significant increase of papers} (\slall) in 2016 and 2017, supporting our assumption that agile for SCS becomes increasingly important.}
While we did find a number of articles in journals and magazines (5) as well as conference papers (22), most of them are quite short.
Together with a considerate number of workshop papers (7
) this suggests that the research field is still young.
In Table \ref{tab:papers}, we also give a rough characterization                                 of the main concern of the selected papers of each year, based on concepts highlighted in the abstracts and keywords.
While the first papers in 2006-2010 were very positive about agile methods, consecutive papers discuss more and more difficulties.
From 2011 on, papers aim to systematically analyse and overcome such obstacles.
We observe proposals to enhance (plan-driven) methods for developing SCS with agile practices, but also suggestions to enhance agile methods with concepts from safety standards.

\begin{table*}[t]
\centering
\caption{Overview of Papers}
\label{tab:papers}
\scriptsize
\begin{tabular}{ccv{.12\textwidth}v{0.22\textwidth}v{0.47\textwidth}}
\toprule
\textbf{Year} & \textbf{Count} & \textbf{Journal} & \textbf{Conference/WS} & \textbf{Topics}\tabularnewline
\midrule
2017 & 5 
& & EuroSPI \citepri{Doss2017}, ICSE \citepri{Vost2017}, Profes \citepri{Wang2017a}, XP \citepri{Hanssen2017}, XP WS \citepri{Wang2017} & Hazard analysis, Safety Assurance, Fitting Safety into Scrum, S-Scrum, Documentation, Safety Story/Epic, Continuous Safety-Builds, Continuous Delivery  \tabularnewline
\grayrow 
2016 & 9  
& Crosstalk \citepri{McMahon2016} &FedCSIS \citepri{Lukasiewicz2016}, ICSSP \citepri{Trektere2016}, RAMS \citepri{Myklebust2016}, SAFECOMP \citepri{staalhane2016agile}, XP \citepri{hanssen2016quality}, SE WS \citepri{wang2016toward}, XP WS \citepri{Doss2016,Stalhane2016} & Agile and CMMI, Agile Principles in Safety Frameworks, Safety Principles in Scrum, Safe Scrum, Early Safety Analysis, Generic Failure Modes, Domain Specific Fault Trees, Compliance, Incremental Design, Safety Case, safety guided design, safety validation\tabularnewline
2015 & 1 & &ICCNEEE \citepri{Abdelaziz2015} & systematic iterative approach, safety modular decomposition, safety argument \tabularnewline
\grayrow 
2014 & 4 & & eChallenges \citepri{Kuchinke2014}, SAFECOMP \citepri{Stalhane2014}, SPICE \citepri{McHugh2014}, XP \citepri{Heeager2014} &  agile (in-process) change impact analysis, hybrid: plan-driven and agile, documentation in agile, quality assurance in agile, compliance by design\tabularnewline
2013 & 4& BI\&T \citepri{Schmidt2013} & ICSE \citepri{fitzgerald2013scaling}, Profes \citepri{notander2013}, SERENE WS \citepri{Gorski2013} & human factors, good documentation, certification cost, component reuse, iterative processes, agile and regulation, conformance \tabularnewline
\grayrow 
2012 & 4 & (CSCI \citepri{gorski2012assessment})$^1$ & SPICE \citepri{McHugh2012}, FormSERA WS \citepri{Wolff2012}, ISSREW WS \citepri{Jonsson2012} & risks of agile, documentation, traceability, regulatory compliance, up front planning, managing multiple releases, formal methods, requirements analysis, change management \tabularnewline
2011 & 2 & IJCCBS \citepri{Paige2011}, \\ SPE \citepri{gary2011agile} & & can agile processes be applicable to SCS? empirical process control, right amount of ceremony\tabularnewline
\grayrow 
2010 & 2 & & AGILE \citepri{ge2010iterative}, LESS \cite{cawley2010lean} & minimal up front design, iterative development, Lean/Agile adoption for SCS, hybrid processes \tabularnewline
2009 & 1 & & AGILE \citepri{Rasmussen2009} & agile is best suitable for FDA regulated medical devices \tabularnewline
\grayrow 
2008 & 1 & & AGILE \citepri{Rottier2008} & test automation, requirements validation, reporting, performance metrics \tabularnewline
2006 & 1 & & XP \citepri{Wils2006} & agile helps with changing requirements of SCS, effectiveness of agile practices changes over time\tabularnewline
\bottomrule
& \multicolumn{4}{l}{ $^1$) Paper added through snowballing since not included in scopus results.}
\end{tabular}
\end{table*}

\subsection{RQ2: Key benefits of applying agile methods to SCS}
With respect to RQ2, we derived key benefits from the papers and grouped them 
into the following categories:
\begin{table*}[t]
\centering
\caption{Solution Candidates in Literature}
\label{tab:solutions:1}
\scriptsize
\begin{tabular}{v{.13\textwidth}v{0.82\textwidth}}
\toprule
\textbf{Solution cand.} & \textbf{Description} \tabularnewline
\midrule
\multicolumn{2}{c}{\emph{Principles}}\tabularnewline \midrule
\emph{Customer involvement at all levels} & \textbf{On-site customer} {\tiny \cite{Jonsson2012}} should be part of hazard analysis, safety analysis, SSRS requirements phase, sprint reviews {\tiny \cite{Doss2016,staalhane2016agile}}
at all levels  of product development {\tiny \cite{Stalhane2016,Doss2017,Schmidt2013}}. 
\textbf{Product owner} can act as the on-site customer {\tiny \cite{fitzgerald2013scaling}}. 
\textbf{Frequent contact with the customer} {\tiny \cite{Heeager2014}} hand  \textbf{regular communication} {\tiny \cite{Wils2006}} allows team to react on the specific needs of the customers.
\tabularnewline
\grayrow \emph{Risk management} & 
Identify high level risks \textbf{in initial phase } {\tiny \cite{Rottier2008}} and consider risks \textbf{during development} {\tiny \cite{Kuchinke2014,Rottier2008}}. Ensure that software is truly safe by employing a \textbf{risk-based approach to planning} {\tiny \cite{Paige2011}}, testing {\tiny \cite{Kuchinke2014}}, verification {\tiny \cite{Paige2011}} and validation {\tiny \cite{fitzgerald2013scaling}}.
\tabularnewline
\midrule
\multicolumn{2}{c}{\emph{{Process and Release planning}}}\tabularnewline \midrule
\emph{Iterative / incremental development} & 
Develop components iteratively {\tiny \cite{ge2010iterative,Paige2011,McHugh2012,hanssen2016quality}} in \textbf{fixed and short iterations} delivering functional software {\tiny \cite{Rasmussen2009,Wils2006}}. Do \textbf{iterative safety analysis} {\tiny \cite{Vost2017,Abdelaziz2015}}, have an \textbf{incremental safety validation plan} {\tiny \cite{staalhane2016agile,Rottier2008}} and \textbf{incremental safety case} {\tiny \cite{staalhane2016agile}}.
\tabularnewline
\grayrow \emph{Backlog management} & 
\textbf{Team should maintain} a groomed, refined, and prioritised backlog {\tiny \cite{Gorski2013,McMahon2016,gary2011agile,staalhane2016agile}} with \textbf{two parts}: one for functional requirements and one for safety requirements {\tiny \cite{Abdelaziz2015,Doss2016,Myklebust2016,wang2016toward,Rottier2008,Wolff2012,hanssen2016quality}}
\tabularnewline
\emph{User stories} & 
Used for product definition {\tiny \cite{Rottier2008,Kuchinke2014}}, high level requirements {\tiny \cite{Hanssen2017}}, upfront planning {\tiny \cite{McHugh2012}}, and \textbf{as basis for safety analysis} {\tiny \cite{Myklebust2016}}. 
Refine with use case diagrams or textual use cases {\tiny \cite{Stalhane2016}}. 
\tabularnewline
\grayrow \emph{Essential upfront plan} & 
Do architectural design and hazard/safety analysis up front {\tiny \cite{Abdelaziz2015,ge2010iterative,Doss2017}}. 
\textbf{Use hazard lists or checklists} from literature {\tiny \cite{Stalhane2016}}; perform analysis of previous failure reports for \textbf{process hazard analysis on the user stories} {\tiny \cite{Stalhane2016}}. Use \textbf{FMEA} {\tiny \cite{notander2013}} for decision support {\tiny \cite{Stalhane2014}}.
\tabularnewline
\emph{Formal change control and prioritisation} & 
Implement formal \textbf{change control process} {\tiny \cite{Rottier2008}} with upfront \textbf{lightweight CIA} {\tiny \cite{Stalhane2014}} to assign quality scores to all requirements {\tiny \cite{Doss2017}} and to prioritize requirements {\tiny \cite{fitzgerald2013scaling,Schmidt2013}}. Maintain \textbf{CIA report} (issues raised and resolved) {\tiny \cite{Stalhane2016,staalhane2016agile}}. \textbf{Formalise design reviews} for validation and verification {\tiny \cite{Rasmussen2009}}\tabularnewline
\midrule
\multicolumn{2}{c}{\emph{{Roles}}}\tabularnewline \midrule
\emph{Self-organising teams} & 
Teams are self-organized and \textbf{empowered} to manage daily tasks of producing software on their own {\tiny \cite{Schmidt2013,McHugh2014}}. Rely on \textbf{collective code ownership} {\tiny \cite{Wils2006}} to allow the whole team to help and keep track of who did what {\tiny \cite{notander2013}}.
\tabularnewline
\grayrow \emph{Expert knowledge } & 
Include \textbf{expert knowledge in team} {\tiny \cite{Gorski2013,gorski2012assessment}}: QA {\tiny \cite{Doss2017}} or \textbf{safety experts} {\tiny \cite{staalhane2016agile,wang2016toward,Wang2017a}} or \textbf{nominated safety team member} {\tiny \cite{Doss2016}}.
\tabularnewline
\midrule
\multicolumn{2}{c}{\emph{{Testing and Continuous Development}}}\tabularnewline \midrule 
\emph{Reviews} & A \textbf{team of peers} with assigned roles performs code reviews {\tiny \cite{gary2011agile,fitzgerald2013scaling}} or more formal technical reviews {\tiny \cite{Abdelaziz2015}}. \tabularnewline
\grayrow \emph{Acceptance and unit testing} & 
Do {unit tests} {\tiny \cite{Wils2006,Rasmussen2009}} and acceptance tests {\tiny \cite{Wils2006,McHugh2014,Paige2011,Gorski2013}} \textbf{once a number of iterations have been completed} {\tiny \cite{McHugh2012}}. 
Apply Test driven development (TDD) to establish high test coverage {\tiny \cite{Jonsson2012,Wils2006,McHugh2014,Hanssen2017,wang2016toward,Wolff2012,hanssen2016quality,cawley2010lean,Rottier2008}} and increased regression testing {\tiny \cite{McMahon2016}}.
Tests should be \textbf{automated} {\tiny \cite{McMahon2016,Rottier2008}}, \textbf{risk-based} {\tiny \cite{Kuchinke2014}}, \textbf{continuous} and extensive {\tiny \cite{gary2011agile,gorski2012assessment}}, and should \textbf{include safety tests} {\tiny \cite{Vost2017}}.
\tabularnewline
\emph{Continuous Build } & 
Apply continuous integration {\tiny \cite{Jonsson2012,Wils2006,Kuchinke2014,Hanssen2017,Paige2011,fitzgerald2013scaling,hanssen2016quality,Schmidt2013,gary2011agile}} supported by automation {\tiny \cite{Rasmussen2009}} and \textbf{include continuous safety builds} {\tiny \cite{Vost2017}}. 
Strive for \textbf{continuous compliance} {\tiny \cite{fitzgerald2013scaling}} in order to facilitate continuous delivery {\tiny \cite{Hanssen2017,fitzgerald2013scaling}}. 
\tabularnewline
\grayrow \emph{Coding standard } & 
Keep high \textbf{coding standards} that all developers should adhere to {\tiny \cite{Jonsson2012,Hanssen2017,Gorski2013}}, supported by \textbf{pair programming} {\tiny \cite{cawley2010lean,Jonsson2012,Wils2006,McHugh2014,Paige2011,Wolff2012}} and refactoring {\tiny \cite{hanssen2016quality,Jonsson2012,Paige2011,Heeager2014}}, e.g. with help of \textbf{refactoring stories} {\tiny \cite{fitzgerald2013scaling}}\tabularnewline
\midrule
\multicolumn{2}{c}{\emph{{Regular meetings}}}\tabularnewline \midrule 
\emph{Daily meetings} & 
\textbf{Daily meetings} {\tiny \cite{fitzgerald2013scaling,hanssen2016quality}} to provide \emph{visibility of safety requirements satisfaction status} {\tiny \cite{Doss2016}} and \emph{promote collective ownership} {\tiny \cite{Rasmussen2009}}. 
\textbf{Daily stand-ups} provide \emph{feedback, communication, and coordination} to manage technical and organizational \emph{dependencies}  {\tiny \cite{Wils2006}}. 
Establish \textbf{weekly goals} and review them at the end of each week during the daily meeting to \emph{focus on team objectives} {\tiny \cite{Rasmussen2009}}.\tabularnewline
\grayrow \emph{Sprint review} & 
\emph{Demonstrate to product owner} what has been done by the team {\tiny \cite{Stalhane2014,Wolff2012,fitzgerald2013scaling,hanssen2016quality,McMahon2016,Wang2017}} and to \emph{create awareness} of activities among team members {\tiny \cite{Wolff2012}}. 
\textbf{Involve all relevant stakeholders} {\tiny \cite{Doss2016}}. 
\textbf{Include hazard analysis} (ibid) and acceptance criteria in review {\tiny \cite{Wang2017,Schmidt2013,McHugh2014}} and implement it as \emph{independent process} for good quality assurance {\tiny \cite{Kuchinke2014}}.\tabularnewline
\emph{Planning meetings and retrospectives } & 
Create \emph{software development plan} {\tiny \cite{Kuchinke2014}} based on features {\tiny \cite{Rasmussen2009}} at the \textbf{planning meeting} {\tiny \cite{Stalhane2014,hanssen2016quality}}. 
Document chosen \emph{modelling approach}, \emph{implementation language}, \emph{development environment} and \emph{assessment tools} { \tiny \cite{Doss2016}} and \emph{determine safety requirements} {\tiny \cite{Wang2017a}}. 
Combine sprint planning with \textbf{retrospective} {\tiny \cite{Stalhane2016,Rasmussen2009,McMahon2016}} to empirically \emph{improve estimations} {\tiny \cite{fitzgerald2013scaling}}.\tabularnewline
\midrule
\multicolumn{2}{c}{\emph{{Safety Engineering Practices}}}\tabularnewline \midrule 
 \emph{Traceability support} &
Use Traceability to support certification as well as claims that requirements have been met {\tiny \cite{McHugh2012}}. 
Use appropriate tools for automation {\tiny \cite{fitzgerald2013scaling}}. 
Use trace information to \textbf{determine which tests need to be rerun} {\tiny \cite{Stalhane2016}}. 
\tabularnewline    
\grayrow \emph{Standard Operating Procedures (SOPs)} &
Rely on SOPs to improve quality management with clear guidelines to developers {\tiny \cite{Kuchinke2014}} and use them as management records for achieving some compliance objectives in agile {\tiny \cite{Hanssen2017}}. 
The records should \textbf{include safety standards the team should be familiar with} {\tiny \cite{Kuchinke2014}} and \textbf{short written policies} for documentation, review and testing, developed by the team {\tiny \cite{Kuchinke2014}}.
\tabularnewline
\bottomrule

\end{tabular}
\end{table*}

\emph{Efficient use of available information:}
{A}gile methods advocate responding to change over following a plan and  {encourage} developers to use available information  to start or continue their work \cite{Stalhane2016}. 
Even early sprints aim to deliver working software {which} helps to identify and {address} technical risks{, e.g. with respect to safety cases,} early \cite{Schmidt2013}.
{This promises to remove the need for} heavy upfront design \cite{wang2016toward}. 

\emph{Improved stakeholder involvement:}
{In SCS development, coordination and understanding between the relevant stakeholders is of utmost importance.}  {Agile methods} enable continuous communication between different stakeholders \cite{Myklebust2016}, thus { allowing them} to contribute to development \cite{Stalhane2016}. 
This keeps the team  better aligned with  stakeholders and ensures that user needs and intended use are well understood \cite{Schmidt2013}{,  increases} customer orientation \cite{Hanssen2017}{,} and  strengthens the trust between stakeholders \cite{Heeager2014}. 
  
\emph{{Improved safety culture:}}
In an agile setting, the available information can be used to get an early start on the safety analysis \cite{Stalhane2016}. 
This helps in identifying possible barriers to safety which also makes it easier to discover and correct faulty system requirements \cite{Myklebust2016}. 
The {practice of }frequent integration {of} software and hardware elements {supports} early identification of {potential} issues \cite{Rasmussen2009}. 
The early focus on safety improves the level of safety awareness within the team and among stakeholders, thus improving the safety culture.

\emph{Improved management of changing requirements:}
Agile methods tend to focus on features during development. 
Together with the iterative nature of agile methods{, this} significantly reduces requirements churn \cite{Rasmussen2009} and provides teams with the ability to manage new {and/}or changing requirements \cite{Hanssen2017}. {While most safety requirements may not change that often, it is important to understand the impact a change, for example in functional requirements, would have on safety.}
This enables teams to re-plan their sprints basing on the most recent understanding of the system under development and {its} requirements \cite{Myklebust2016}.

\emph{{Improved} prioritisation:} 
{Agile methods} generally give the advantage of better prioritisation during the handling or managing of changing requirements \cite{Stalhane2014}. 
With prioritisation, focus is put on the highest value feature first, which leads 
{to more thorough definition and validation of the important requirements} \cite{Schmidt2013}.
This allows to put safety related features into the center of attention when developing SCS. 

\emph{Mapping of functional and safety requirements:} 
Studies propose maintaining {specific} backlogs {for} safety and functional requirements such that the development takes safety into context from the start. 
This practice supports a mapping between the functional and safety requirements \cite{Myklebust2016} {and} seamless integration of safety and software engineering \cite{Vost2017}.
    
\emph{Reduced costs:}
Agile methods encourage simple designs which in turn produce simpler software {with reduced} development and maintenance cost \cite{Myklebust2016}. The high response to requirements' changes that agile methods advocate helps to reduce the amount of rework \cite{McHugh2014}{ and} shortens {overall time} spent on development {as well as} the cost that would be spent on redoing the work \cite{Rasmussen2009,Hanssen2017,gorski2012assessment}. 
 {High development costs are a big challenge in developing SCS and the ability to reduce cost and lead time for bringing new safety-critical functions on the market will allow to increase the speed at which organizations learn on how to make such functions even safer. }

\emph{Better test cases:}
The iterative development allows the team to only prepare {and maintain} documents that are needed either for development or certification \cite{Myklebust2016}. 
This also produces {a lower number of tests, but with better focus on their intent,} making it easier to understand and debug \cite{Rasmussen2009}.

\emph{Improved quality:}
SCS development is a quality matter which is usually checked through rigour in testing. 
In agile methods, the continuous availability of working software and early response to changes facilitates ongoing testing, which reduces \ risk to product quality \cite{Rasmussen2009,Schmidt2013,gorski2012assessment}. 
Rottier {and Rodrigues} report that the use of iterative development together with Test Driven Development (TDD) produced fairly consistent high quality code for their product \cite{Rottier2008}.

\emph{Improved opportunities for reuse:}
Good adaption of agile methods helps build{ing} new frameworks that can be reused {in later phases or projects, which will improve efficiency} \cite{Myklebust2016,Rottier2008}. 
{Such} frameworks can also enable the reuse of test cases 
and reduce the overall workload \cite{Wolff2012}.

\subsection{RQ3: Challenges with agile development of SCS}
{Several of our selected papers identified challenges relating to use of agile methods in SCS development, which we group and summarize into the following themes:} 

\emph{Difficult to manage knowledge flow between many stakeholders:}
There are many stakeholders in SCS development, including other engineering disciplines, sub-system suppliers, users, {and} external bodies such as certification authorities \cite{Paige2011,gorski2012assessment}. 
{Managing communication among these stakeholders is challenging.
While agile methods may improve stakeholder involvement (see above), stakeholders of SCS} are used to communication {through} documents. 
{The lack of focus on documentation in agile methods may lead to} unclear safety documentation{, thus hindering} effective communication between the different stakeholders \cite{Wang2017a,Wang2017}.

\emph{Safety standards facilitate a waterfall mindset:}
Safety standards {are} prescriptive and {often described with} a waterfall-based process {in mind} \cite{notander2013}.
{With this mindset, regulated domains often value effectiveness of waterfall processes over the flexibility of agile methods} \cite{fitzgerald2013scaling}. 
For example, many safety standards demand that one person must not create and review the same artifact \cite{notander2013,Jonsson2012,Kuchinke2014}, while agile methods favor cross-functional teams.
This {mismatch of the mindset} causes risks to potential compliance \cite{Schmidt2013}. 

The waterfall mindset also shows a \emph{strong focus on documentation}. 
In SCS, documentation is required for {showing} regulation compliance \cite{McHugh2012} {and is the preferred way of} communication with certification bodies \cite{Paige2011}. 
Documentation is also the primary evidence for traceability \cite{fitzgerald2013scaling}. 
{Agile methods are often reported to suffer from this focus on documentation}  \cite{Wang2017,Kuchinke2014,fitzgerald2013scaling,Paige2011,Trektere2016,Heeager2014} as extensive documentation will diminish advantages of agility while lack of documentation will lead to insufficient traceability}.

\emph{Lack of trust in agile methods:}
Since agile methods do not give clear guidelines for project monitoring and control, it is hard to determine the sufficient level of evidence to present \cite{Doss2016} for certification. 
Thus, it appears difficult to estimate effort and through the ``unstructured nature'' of agile methods, the amount of effort for delivering a feature seems to be underestimated \cite{Rottier2008}.
The maturity of  agile methods is hard to compare to CMM and ISO standards \cite{Paige2011} and many managers are not convinced about business benefits of agile methods \cite{fitzgerald2013scaling}. 
There is also doubt about whether agile methods can provide sufficiently rigorous testing \cite{Paige2011}.

\emph{Upfront planning:} 
Safety standards suggest upfront design, hazard identification, and analysis. 
With agile methods, it is hard to determine how much time is enough for upfront planning \cite{McHugh2012,Wang2017a}
and consequently difficult to  identify all of the ways in which software can contribute to system level hazards up front \cite{Doss2016}.
Thus, there is the perception that planning in agile methods is insufficient for SCS development \cite{Wang2017,fitzgerald2013scaling}. 
The aim for short upfront design and for analysing requirements just in time during iterations puts time pressure for determining the safety requirements and makes it difficult to evaluate the quality of the safety arguments \cite{ge2010iterative,Wang2017a} which could impede certification \cite{gorski2012assessment}. 
Even more, attempts to include safety in agile can shift focus from customer value towards verification and validation efforts \cite{staalhane2016agile}.

\emph{Flexibility vs. Safety:}
Agile methods do not provide practical guidelines for change impact analysis \cite{Stalhane2014}, yet they provide flexibility in managing requirements. 
Every update therefore calls for relentless testing and strict configuration management \cite{Hanssen2017} and un-coordinated software changes can lead to increased complexity as the project progresses \cite{Wils2006}.
In addition, it is difficult to manage requirements that change iteratively \cite{Heeager2014}.

\subsection{RQ4: Solution candidates (e.g. principles and practices) for challenges with respect to agile development of SCS}}
We present solutions collected from the selected papers in Table \ref{tab:solutions:1}. 
Some of these solutions relate to enhancing agile, others to complementing safety, which emphasizes the aim of these works to bring both worlds closer together.
We organized the proposed solutions in different categories, starting with \emph{principles} that were suggested to provide guidance for agile development of SCS.
These are rather abstract guidelines {or} goals to strive for when implementing an agile way of working with SCS.
In addition, we identify proposed solutions with respect to  \emph{process and release planning} and \emph{roles}.
%
We then continue with more concrete practices that were proposed in literature, which we grouped in relation to \emph{testing and continuous development}, \emph{regular meetings}, and broader \emph{safety engineering practices}.

\subsection{{Synthesis of Findings}}
\begin{figure}[b]
\centering
\includegraphics[width=0.4\textwidth]{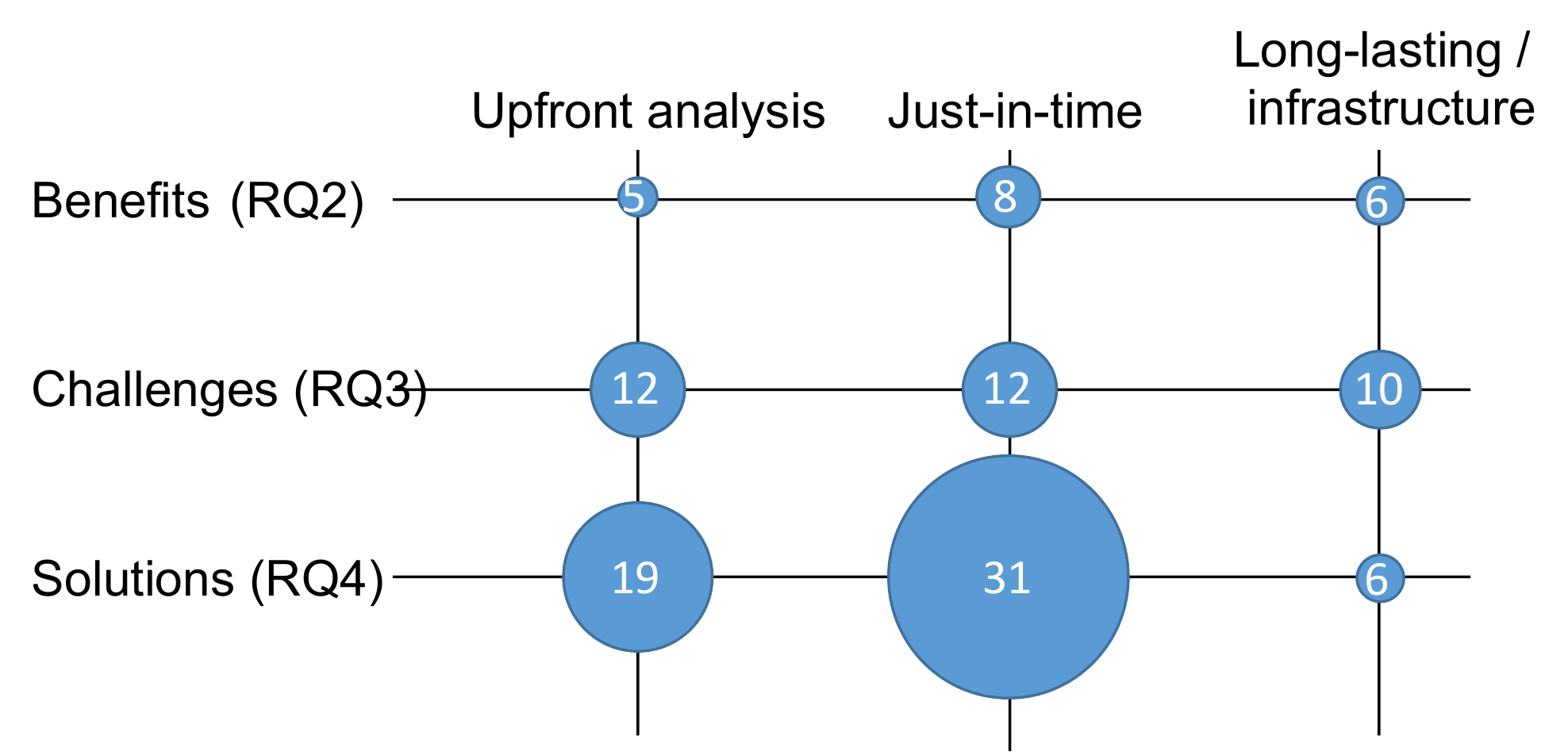}
\caption{Bubble chart with numbers of papers per category.}
\label{fig:bubble-chart}
\end{figure}

{In Table \ref{tab:lit}} we give an overview of key benefits, challenges, principles, and practices discussed in literature and in Fig. \ref{fig:bubble-chart} we show how the selected papers are distributed over these categories. 
\begin{table}[h]
\caption{Summary of literature}
\label{tab:lit}
\centering
\resizebox{\columnwidth}{!}{%
\begin{tabular}{lr}
\toprule
\grayrow  \textbf{Benefits}&Total Pri. Papers \tabularnewline
  \midrule
Improved stakeholder involvement &5\tabularnewline
 Reduced costs &5\tabularnewline
 Improved quality & 4\tabularnewline
 Efficient Use of available information&3\tabularnewline
 Improved safety culture &3\tabularnewline
 Improved opportunities for reuse & 3\tabularnewline
 Improved management of changing requirements & 3\tabularnewline
 Improved prioritisation &2\tabularnewline
 Mapping of functional and safety requirements &2\tabularnewline
 Better test cases & 2\tabularnewline
 
 \grayrow\textbf{Challenges}&\tabularnewline
 \midrule
 Strong focus on documentation & 9\tabularnewline
 Upfront planning &9\tabularnewline
 Safety standards to facilitate a waterfall mindset & 6\tabularnewline
 Lack of trust in agile methods &5\tabularnewline
 Flexibility vs Safety &4 \tabularnewline
 Difficult to manage knowledge flow between stakeholders & 4\tabularnewline
\grayrow \textbf{Practices and principles}&\tabularnewline \midrule
 Acceptance and unit testing &18 \tabularnewline
 Coding standard&14\tabularnewline
  Continuous build&11 \tabularnewline
 Backlog management& 11\tabularnewline
Iterative development& 10\tabularnewline
Sprint review &10\tabularnewline
 Planning meetings and retrospectives & 9\tabularnewline
 Formal change control and prioritisation& 8\tabularnewline
 Expert knowledge& 7\tabularnewline
  Customer involvement at all levels&7 \tabularnewline
 User stories & 6\tabularnewline
 Essential upfront plan& 6\tabularnewline
 daily meetings&5\tabularnewline
  Self-organising teams & 4\tabularnewline
  Traceability support & 3\tabularnewline
  Risk management & 3\tabularnewline
 Reviews& 2\tabularnewline
 Standard operating procedures & 2\tabularnewline\bottomrule
\end{tabular}}
\end{table}
The majority of work relates to just-in-time activities, such as analysis and management of functional or system safety.
Fewer selected papers relate to upfront analysis, which in many publications is taken for granted.
The smallest amount of papers relates to long-term aspects and infrastructure that lasts beyond an individual project.

It is noteworthy that solution candidates dominate for upfront and just-in-time aspects.
In contrast, we found relatively fewer papers relating to solutions with respect to long-term and infrastructure aspects,  especially in relation to the amount of challenges found in this category.
We found only few papers that discuss a trade-off between upfront analysis and just-in-time analysis or attempt to push more activities from the upfront phases into continuous development. 

This result resonates well with the discussions during our initial workshop. 
Our industry participants highlighted the following challenges that we did not yet found sufficiently supported in the selected papers.

\emph{Upfront:} Our industry participants saw \emph{benefits} in upfront analysis for agile development of SCS, e.g. with respect to reducing dependencies and confining safety-critical or secret software (e.g. due to IP constraints). 
\emph{Challenges} relate to a lack of experience with continuous development of existing products in contrast to developing new products.

\emph{Just-in-time:} Industry participants saw \emph{benefits} with respect to managing continuously changing dependencies and requirements. 
The ability to do continuous assessment and certification was seen as a real enabler.
Also, agile documentation, i.e. focusing on the essential, product-related information was seen as a potential benefit. 
\emph{Challenges} relate to constructing the big picture from local information, quality assurance and assessors with waterfall mindset, and  the question whether product or process based evidence for safety would be the better choice. 
With respect to \emph{solutions}, they showed big hopes in applying specification by example and hardening sprints.

\emph{Long-term/infrastructure:} In this category, our participants saw the biggest need for new concepts. 
Potential \emph{benefits} could include investment in infrastructure with long-term benefits, organization-level assessment, tools to check dependencies, and semi-automation for traceability.
The biggest \emph{challenges} were seen in a pay-per-product mentality, lack of traceability, too many layers of requirements (making it impossible to keep safety requirements up-to-date in continuous development), certification of tools (which will slow down their evolution), old assessment frameworks, and lack of trust in agile maturity.

\section{Discussion and Conclusion}
In this paper, we provide a systematic mapping of agile development of SCS. 
We contribute an overview of research papers discussing experience from industry from 2001 to 2017 in the hope that this will enable future research in this area. 
Our synthesis is based on iterative analysis of selected papers and discussions in a workshop with experts from six major Swedish product companies. 
We found that potential benefits, challenges, and proposed solutions relate to upfront, just-in-time, or long-term and infrastructural aspects.
While companies continue to struggle in all three areas, we see a need for future research specifically in two areas:
a) Investigation of trade-offs between effort done upfront and just-in-time together with guidelines on how to shift more effort into just-in-time analysis; b) Investigation on how to establish beneficial infrastructure for long-term support.

\textbf{Acknowledgements:} We thank all participants in our workshop for their great support, deep discussions, and clarifications throughout this work whenever needed. This work was supported by Software Center (\url{www.software-center.se}) and the SIDA BRIGHT project.

\bibliographystylepri{alpha}
\bibliographystylesec{abbrv}
\bibliographypri{SafetyRefs.bib}
\bibliographysec{RashidahRefs.bib,xp-safety-refs.bib}
\end{document}